\documentclass[preprint]{aastex631}

\newcommand{\add}[1]{{#1}}
\newcommand{\rev}[1]{{#1}}
\newcommand{\dbob}{\ensuremath{b/B_0}}
 \usepackage{graphicx}
 \usepackage{amsmath}
\usepackage{booktabs}

\shorttitle{Variation in Path Lengths}
\shortauthors{Sonsrettee et al.}
\begin{document}
\title{Variation in Path Lengths of Turbulent Magnetic Field Lines and Solar Energetic Particles}
\author[0000-0001-9105-8918]{Wirin Sonsrettee}
\affiliation{Faculty of Engineering and Technology, Panyapiwat Institute of Management, Nonthaburi 11120, Thailand}

\author{Piyanate Chuychai}
\affiliation{33/5 Moo 16, Tambon Bandu, Muang District, Chiang Rai 57100, Thailand}

\author[0000-0002-3764-8949]{Achara Seripienlert}
\affiliation{National Astronomical Research Institute of Thailand (NARIT), Chiang Mai, 50180, Thailand}

\author{Paisan Tooprakai}
\affiliation{Department of Physics, Faculty of Science, Chulalongkorn University, Bangkok 10330, Thailand}

\author[0000-0001-7771-4341]{Alejandro S\'{a}iz}
\affiliation{Department of Physics, Faculty of Science, Mahidol University, Bangkok 10400, Thailand}

\author[0000-0003-3414-9666]{David Ruffolo}
\correspondingauthor{David Ruffolo}
\email{david.ruf@mahidol.ac.th}
\affiliation{Department of Physics, Faculty of Science, Mahidol University, Bangkok 10400, Thailand}

\author[0000-0001-7224-6024]{William H.~Matthaeus}
\affiliation{Department of Physics and Astronomy, University of Delaware, Newark, DE 19716, USA}
\affiliation{Bartol Research Institute, University of Delaware, Newark, DE 19716, USA}

\author[0000-0002-7174-6948]{Rohit Chhiber}
\affiliation{Department of Physics and Astronomy, University of Delaware, Newark, DE 19716, USA}
\affiliation{Heliophysics Science Division, NASA Goddard Space Flight Center, Greenbelt, MD 20771, USA} 

\begin{abstract}
Modeling of time profiles of solar energetic particle (SEP) observations often considers transport along a large-scale magnetic field with a fixed path length from the source to the observer. Here we point out that variability in the turbulent field line path length can affect the fits to SEP data and the inferred mean free path and injection profile. To explore such variability, we perform Monte Carlo simulations in representations of homogeneous 2D MHD + slab turbulence adapted to spherical geometry and trace trajectories of field lines and full particle orbits, considering proton injection from a narrow or wide angular region near the Sun, corresponding to an impulsive or gradual solar event, respectively. 
We analyze our simulation results in terms of field line and particle path length statistics for $1^\circ\times 1^\circ$ pixels in heliolatitude and heliolongitude at 0.35 and 1 AU from the Sun, for different values of the turbulence amplitude $b/B_0$ and turbulence geometry as expressed by the slab fraction $f_s$. 
Maps of the most probable path lengths of field lines and particles at each pixel exhibit systematic patterns that reflect the fluctuation amplitudes experienced by the field lines, which in turn relate to the local topology of 2D turbulence. 
We describe the effects of such path length variations on SEP time profiles, both in terms of path length variability at specific locations and motion of the observer with respect to turbulence topology during the course of the observations.\footnote{Accepted for publication in the {\it Astrophysical Journal}.}
\end{abstract}

\section{Introduction}

Due to the impact of space weather effects from solar storms on human activity, there is a significant community effort focused on analyzing historical solar events and predicting solar energetic particle (SEP) fluxes \cite[see][and references therein]{EngelbrechtEA22}.
The interplanetary transport of SEPs, including scattering due to magnetic fluctuations in the solar wind, typically plays a dominant role in determining their intensity and anisotropy profiles \citep{MeyerEA56,Jokipii66,Earl76b,Ruffolo95,BieberEA02},
which are a key aspect of space weather effects of solar storms \citep{Knipp11}.

\begin{figure}
\includegraphics[width=.5\linewidth]{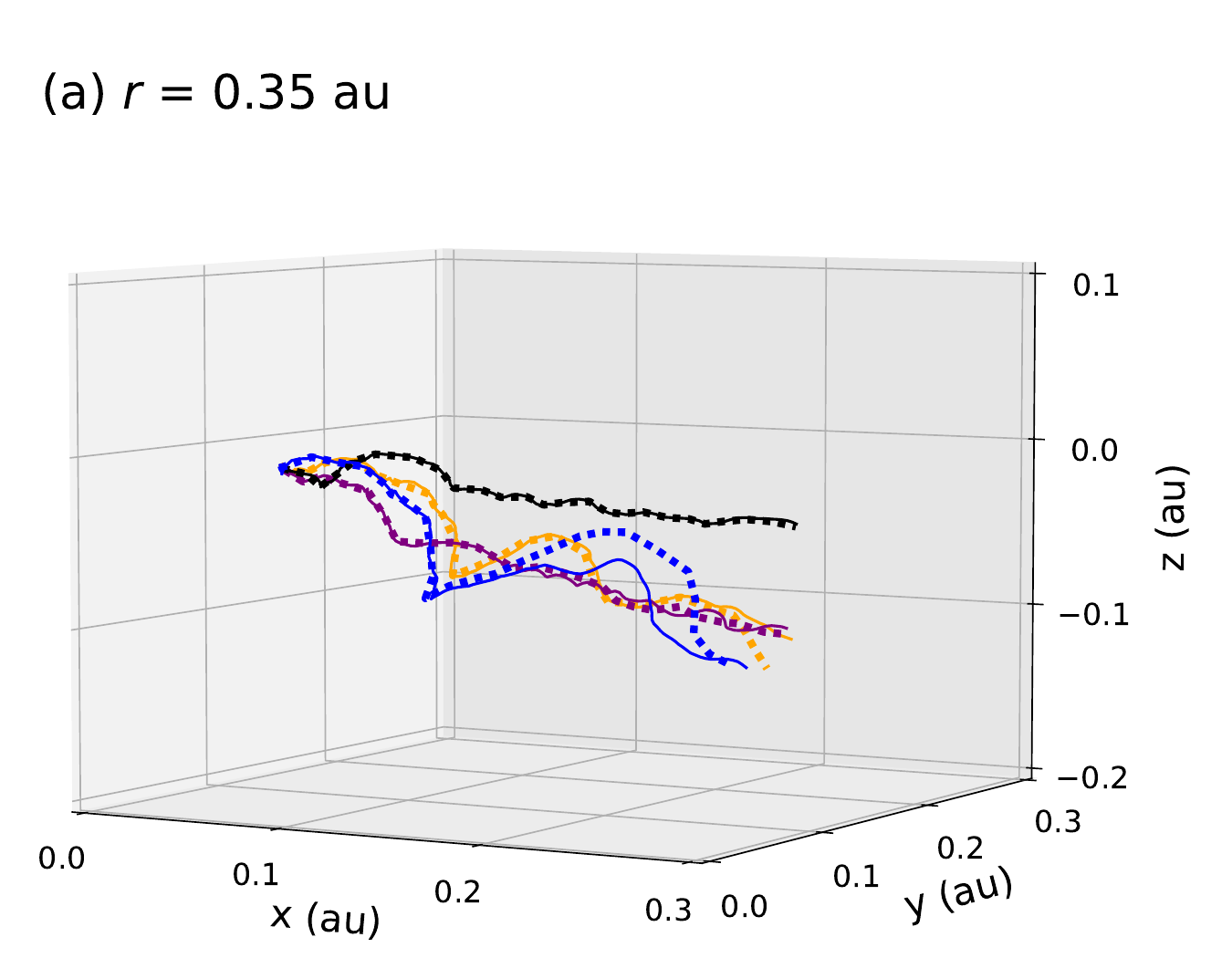}
\includegraphics[width=.5\linewidth]{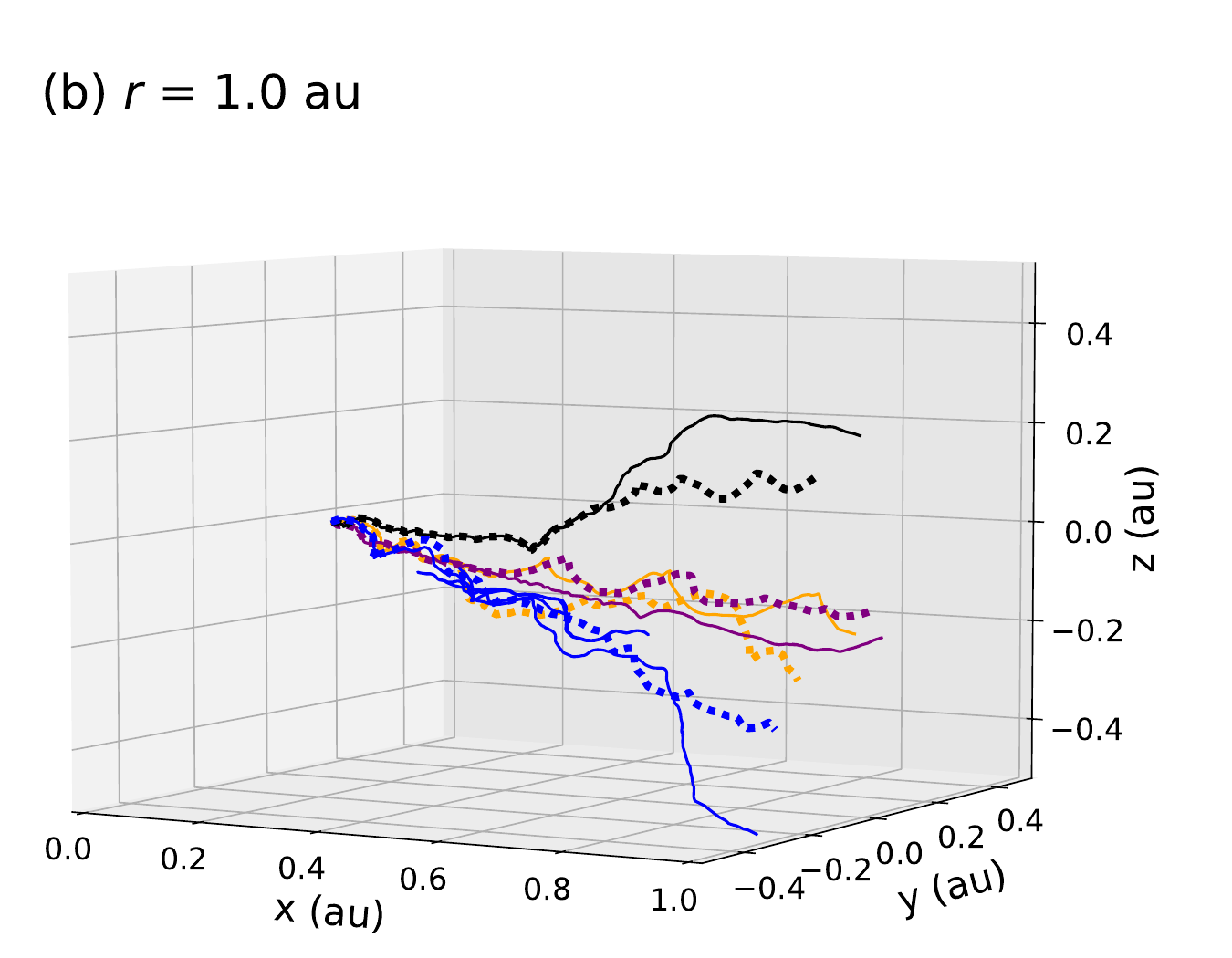}
\caption{
Illustration of  sample 1-MeV proton trajectories (solid lines) and their associated field line trajectories (dashed lines) from the same initial location, based on full-orbit simulations in a representation of the 2D+slab model interplanetary turbulence with normalized rms magnetic fluctuation $b/B_0=1$ and slab fraction $f_s=0.2$.  
(a) The trajectories are traced to $r = 0.35$ au from the Sun. We can see that the particles mostly closely follow the field lines.   
(b) But when we extend them to $r = 1.0$ au, more deviation between particle and field line trajectories can be observed.}
\label{fig:orientation}
\end{figure}

SEP trajectories can be considered to be determined by those of the magnetic field lines combined with the effect of parallel scattering \citep{ MatthaeusEA03,ShalchiEA04,MinnieEA09}. 
Figure \ref{fig:orientation} provides an illustration of this point based on full-orbit simulations of sample particle and field line trajectories in a static representation of interplanetary turbulence. 
Globally, particles do not strictly follow their initial field line, even to 0.35 au.
When extended to $r = 1$ au, a significant deviation between particle and field line trajectories can be observed, including one trajectory (dark blue) that moves outward, backscatters, and then moves outward again along a slightly different path.
 Nevertheless, statistically the trajectories are {\it locally} related to field line trajectories, with occasional backscattering due to magnetic turbulence.    

Efforts toward prediction and modeling of SEP arrival at the Earth and other locations in the heliosphere \citep[see][and references therein]{vandenBergEA20} relate closely to the distribution of SEP path lengths, which in turn relate to path lengths of magnetic field lines as noted above.
The full-orbit path length $s$ of particles directly relates to the time traveled since injection as $\Delta t=s/v$, especially if we neglect the momentum changes due to adiabatic deceleration over this short time scale \citep{Ruffolo95} and assume that a particle travels at a constant speed $v$.  
Then the path length distribution represents the time-intensity profile that would result from an instantaneous SEP injection near the Sun.
To account for a realistic, non-instantaneous injection of SEPs, some models derive {\it ab initio} expectations of particle injection and perform forward modeling of their transport, while others perform backward modeling to fit observed data, deconvoluting the effects of transport to infer a scattering mean free path \citep[see results summarized by][]{EngelbrechtEA22} and/or SEP injection profiles.  
Some models require an input value for the path length of the magnetic field line along which particles are being transported.

Observationally, velocity dispersion analysis
\citep{LinEA81,TylkaEA03}
has been applied to fit the onset times $t$ for various particle energy channels, each with a corresponding particle velocity $v$, which are fit to an equation
\begin{equation}
    t=t_{\mathrm{start}}+s/v,
\end{equation} 
where $t_{\mathrm{start}}$ is interpreted as the start time of injection.
A common interpretation is that scattering effects can be neglected for first-arriving particles, so $s$ is interpreted as the particle path length along the large-scale guiding magnetic field.
However, it has been pointed out that more generally, if scattering is not neglected, $s$ should be interpreted as the path length of initially arriving particle orbits, including their gyration around the magnetic field.
Various studies have suggested that the scattering effects should not be neglected \citep{LintunenVainio04,SaizEA05},
including a recent analysis of SEP ions observed using the {\it Parker Solar Probe} (PSP), from which $s\approx0.625$ au was inferred at a radial distance of $r\approx0.35$ au from the Sun \citep{ChhiberEA21aa}. 
[In another analysis of the same event, \citet{ChengEA23} obtained an even larger value of $s\approx0.70$ au.]
In this case it seems particularly unlikely that $s$ represents the path length of the large-scale interplanetary magnetic field; rather the difference between $s=0.625$ and $r=0.35$ was attributed to a combination of the enhanced path length of random-walking magnetic fields and pitch angle scattering, leading to an effective pitch angle of 25$^\circ$ for ions arriving at the observed onset time.

\citet{ChhiberEA21aa} provided both simple and rigorous estimates of field line path lengths as a function of relative turbulence amplitude, and tested them using Monte Carlo simulations of field line trajectories. 
Based on full-orbit simulations of proton trajectories in a representation of interplanetary turbulence, they found that the mean path length of guiding center trajectories of energetic particles was slightly shorter than that of the random-walking field lines, but full-orbit path lengths $s$ were significantly longer, because the pitch angle is non-zero due to interplanetary scattering.  
In particular, near the Sun the magnetic field is nearly radial and the strength of adiabatic focusing is proportional to $r^{-1}$, and this process drives the particle pitch angle toward zero, but farther from the Sun as this process weakens, pitch angle scattering drives the \add{distribution to larger values of the pitch angle} \citep{RuffoloK95}.  
Simulated path length distributions have also been presented for field lines by \citet{LaitinenEA23a} and \citet{LiBian23} and for energetic particles by \citet{LaitinenEA23b}.
\add{Various studies recognize that while nominal path lengths are computed relative to a typical Parker spiral magnetic field \citep{Parker58},}  
it is worth noting that inside an interplanetary magnetic flux rope (e.g., inside an interplanetary coronal mass ejection), path lengths of field lines and particles can be much longer \citep[e.g.,][]{LarsonEA97,WimmerEA23}.

Here we extend 
%that work
the work of \citet{ChhiberEA21aa} 
to examine distributions and spatial variations of field line and particle path lengths
and their dependence on selected turbulence parameters.
We trace field line and particle trajectories from either a narrow SEP injection region near the Sun, corresponding to an  impulsive solar event, or a wide injection region, corresponding to a gradual solar event \citep{Reames99}. 
\add{For} an instantaneous injection near the Sun, \add{the} particle path length distribution 
corresponds to a time-intensity profile of SEPs
with the pulse and wake that are characteristic of focused transport \citep{Earl76b}, though it is a global distribution in the sense of integrating particle crossings over a sphere at distance $r$.
To examine spatial variations, we divide heliolongitudes and heliolatitudes into $1^\circ\times1^\circ$ ``pixels,'' allowing us to construct maps of the most probable (``peak'') path length of field lines or particles arriving at each \add{pixel.}
We find systematic spatial variations in peak path lengths that relate to the strength and topology of the solar wind magnetic turbulence.
We also discuss implications of our results for SEP transport modeling.

\section{Methods}
\label{sec:meth}

Our simulation of magnetic turbulence in the solar wind is based on a idealized two-component fluctuation model with slab and 2D components, motivated by observations of a ``Maltese cross'' pattern of solar wind magnetic fluctuations varying predominantly in directions parallel or perpendicular to the mean field, respectively \citep{MatthaeusEA90,DassoEA05}.
This model has been found to provide a useful description of observed solar wind fluctuations \citep{BieberEA96} and energetic particle transport \citep{BieberEA94,EngelbrechtEA22}.
We emphasize that this two-component model is a kinematic representation and is not intended as a dynamical model of turbulence.
As expressed in spherical geometry, the 2D+slab model would ideally have the form \citep{RuffoloEA13,TooprakaiEA16}:
 \begin{equation}
 \mathbf B\left(\mathbf r \right) = \mathbf B_0 \left(\mathbf r \right) + \mathbf{b}(\mathbf r)
 = \frac{r_{1}^{2}}{r^{2}} [B_1\hat{\mathbf r} + \mathbf{b}^\mathrm{slab}\left ( r \right )+\mathbf{b}^\mathrm{2D}\left ( \mathbf \varphi, \Lambda  \right ) ].\hfill
 \label{eq:1}    
 \end{equation}

A radial mean field was used in order to allow the construction of a 2D fluctuation field, $b^\mathrm{2D}(\varphi,\Lambda)$, that depends only on the heliolongitude $\varphi$ and heliolatitude $\Lambda$ 
\add{and is statistically homogeneous in those coordinates}.  

The \rev{mean} magnetic field \rev{$\mathbf{B_0}$} is proportional to $r^{-2}$ \rev{and its}
magnitude is $B_1 =$ 5 nT at $r_1 =$ 1 au. The two components of the fluctuation model are the slab fluctuation $\mathbf{b}^\mathrm{slab}$ and the 2D fluctuation $\mathbf{b}^\mathrm{2D}$, which are directed perpendicular to the mean field. 
\rev{We use the notation $b$ to refer to the rms magnetic fluctuation, and for}
simplicity \rev{$b/B_0$} is taken to be independent of $r$, although PSP observations  indicate that it actually varies as $r^{1/4}$ inside Earth\rev{'s} orbit \citep{Chhiber22}.  

The 2D component $\mathbf{b}^\mathrm{2D}$ depends on the perpendicular coordinates: $\varphi$ and $\Lambda$. 
We can define a magnetic potential $a(\varphi,\Lambda)$ so that 
\begin{equation}
\mathbf{b}^\mathrm{2D}(\varphi,\Lambda) = \nabla\times[a(\varphi,\Lambda)\mathbf{\hat r}].
\end{equation}
Then $\mathbf{b}^\mathrm{2D}$ is perpendicular to the gradient of $a$ and follows equipotential contours in $\varphi$ and $\Lambda$, causing the topological trapping of field lines identified as an explanation for SEP dropout observations \citep{RuffoloEA03}.
In order to generate more realistic 2D fluctuations with coherent structures, we use a 2D magnetohydrodynamic (MHD)  simulation \citep{TooprakaiEA16}. The simulation started with an initial state corresponding to a Kolmogorov power spectrum of turbulence, and then a numerical code was used to evolve the field according to 2D MHD \citep{SeripienlertEA10}.
The 2D magnetic potential \add{at $1024\times1024$ spatial grid points} is then mapped from Cartesian coordinates $(x,y)$ to angular coordinates $(\varphi,\Lambda)$ \add{\citep{RuffoloEA13}}, with the total 2D correlation scale \citep{MatthaeusEA07} set to $\lambda_{c2}=0.0123$ au \add{at $r=1$ au}, similar to the value measured by \citet{WeygandEA09},
neglecting the mild \add{radial} variation inferred from PSP data \citep{CuestaEA22}.

Similarly, the slab fluctuation field $\mathbf{b}^\mathrm{slab}$ is generated along one dimension, $z$, from a turbulence power spectrum with the \add{constant} bendover scale \add{$\lambda=0.02$} au \add{(corresponding to a correlation length of 0.0149 au)} and an inertial range with a Kolmogorov spectrum \citep{TooprakaiEA16}.
\add{
We use a fast Fourier transform, with that turbulent power spectrum over regularly spaced wavenumbers from $2\pi/L$ to $N\pi/(2L)$ and zero power (``zero padding'') at higher wavenumbers,
where $N=2^{22}=4,194,304$ is the number of grid points and $L$ is the length of the periodic box. 
We set $L=10,000\,\lambda$ to avoid periodicity effects, which are particularly important for a fluctuation field that varies in only one dimension.
} 
When mapping onto angular coordinates, in order to maintain $\nabla\cdot\mathbf{b}^\mathrm{slab}=0$, we need to modify Equation (\ref{eq:1}) to use $b_\varphi^\mathrm{slab}(r)=b_x^\mathrm{slab}(z)$ and $b_\Lambda^\mathrm{slab}(r,\Lambda)=b_y^\mathrm{slab}(z)\sec\Lambda$.
To some degree, this violates homogeneity, but we use a simulation domain of $-25^\circ\leq\varphi\leq25^\circ$ and $-25^\circ\leq\Lambda\leq25^\circ$,
\add{which corresponds to the periodic box size of the 2D MHD simulation}, so the factor of $\sec\Lambda$ has only a minor effect.

To investigate the variations of path lengths of solar energetic particles (SEPs), we use a Monte
Carlo approach, tracing field lines and particles from either a narrow or wide injection region at $r_0=0.1$ au, corresponding to an impulsive or gradual solar particle event, respectively. 
Because strong adiabatic focusing (i.e., magnetic mirroring) close to the Sun should rapidly reduce the particle pitch angle \citep{RuffoloK95}, we inject particles with an initial pitch angle of zero, i.e., with a velocity parallel to the local $\mathbf{B}$.
For a narrow injection,  
we randomly inject magnetic field lines and particles within a $5^\circ$-diameter circle 
centered at $\varphi=\Lambda=0$
\citep{TooprakaiEA16}. 
In this case we do not use periodic boundary conditions in $\varphi$ and $\Lambda$, and trajectory tracing is terminated if a field line or particle reaches those boundaries.
To model a wide injection,
we use uniformly random injection positions over the entire $(\varphi,\Lambda)$ domain, with periodic boundary conditions in $\varphi$ and $\Lambda$ to efficiently model homogeneous injection from a wider range of angles \citep{RuffoloEA13}. 
\rev{Periodic boundary conditions imply that a trajectory that leaves one edge of the domain is replaced by one entering at the opposite side, so that the injection region is effectively repeated beyond the original domain.}

The streamline equation  \begin{equation}
 \frac{\mathrm{d}r}{B_r} = \frac{r \,\mathrm{d}\Lambda}{B_\Lambda} = \frac{r \cos \Lambda \,\mathrm{d} \varphi}{B_\varphi}.
 \end{equation}
is used to trace the trajectories of magnetic field lines in spherical geometry.
 For particle trajectories, we solve the Newton-Lorentz equation
\begin{equation}
\gamma m \frac{\mathrm{d}\mathbf{v}}{\mathrm{d}t} = q\mathbf{v}\times\mathbf{B}(\mathbf{r})
\end{equation}
to update the particle position $\mathbf{r}$ and velocity $\mathbf{v}$, where $m$ and $q$ are the particle mass and charge, respectively.
 Both field line and particle trajectories are traced numerically using a version of the {\sc streamline} code \citep{DalenaEA12}.
The path length of a field line is evaluated as 
\begin{equation}
S=S_0+\int_{r_0}^{r}(\mathrm{d}S/\mathrm{d}r)\mathrm{d}r
\label{eq:S}
\end{equation}
\citep{ChhiberEA21aa}, where $S_0$ represents the path length inside $r_0$, which we simply set to $r_0$ so that a radial field line with no turbulence has $S=r$.  
An analogous formula is used to calculate the path length $s$ of a particle.
Since we assume transverse fluctuations, for field lines we have
\begin{equation}
\mathrm{d}S/\mathrm{d}r=\sqrt{1+b^2(\mathbf{r})/B_0^2}.
\label{eq:dSdr}
\end{equation}

\begin{figure}
\centering
\includegraphics[width=.6\linewidth]{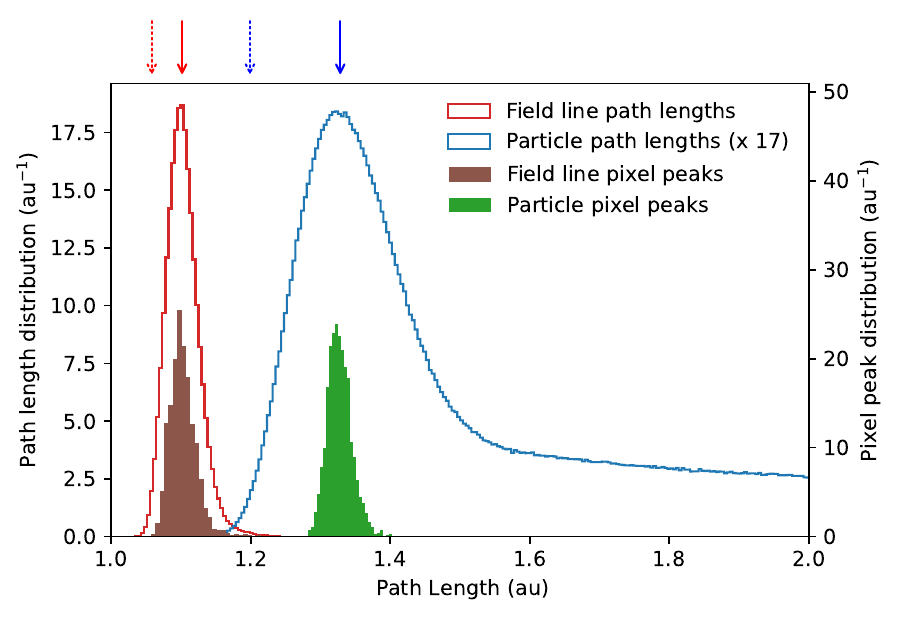}
\caption{Path length distributions at $r=1$ au from a simulation of 5 million sets of field lines (red) and 1-MeV protons (blue) for a wide injection region (as for a gradual SEP event) with $b/B_0=0.5$ and $f_s=0.2$. 
\add{Solid arrows indicate most} probable (``peak'') path lengths 
\add{and dashed arrows indicate the path lengths 
where the 
distribution reaches 10\% of its maximum.}
For an instantaneous particle injection, the particle path length distribution corresponds to a time-intensity profile.
Distributions of peak path lengths within ``pixels'' of $1^\circ\times 1^\circ$ in heliolatitude and heliolongitude
for the field lines (brown) and particles (green) reflects the spatial variation of peak path lengths.}
\label{fig:pathlength_dist}
\end{figure}

We record and analyze all crossings of field lines and particles at the spherical shells at distances $r_A=0.35$ and $r_B=1.0$ au from the Sun.  \rev{Particle trajectories are traced for times $t$ such that
$0\leq vt\leq17.4$ au.}
\add{
Figure \ref{fig:pathlength_dist} shows an example of 
the global distributions of 
field lines (red) and particles (blue), to all heliolongitudes and heliolatitudes at the distance $r$ of interest.}

To examine the spatial variability of path lengths, we divide the domain of heliolongitude and heliolatitude into 2,500 square-degree boxes, called ``pixels.''
\rev{For distributions of particle path lengths, when interplanetary scattering plays a major role, standard measures such as the mean or median become sensitive to the duration of the simulation and may grow without bound as the duration increases; we therefore choose to characterize path length distributions by a most probable or ``peak'' path length.}
We have developed an algorithm to fit path length distributions, possibly with as few as 70 path length values, to estimate the most probable (``peak'') value. 
\add{The following description is for the peak particle path length $s_p$, and it also applies to the peak field line path length $S_p$.}
Since we \add{want to include} the contribution of \add{backscattered} particles, \add{we include multiple crossings of} each individual Monte Carlo particle or field line \add{across} the spherical shell \add{at radius $r$}; however, we limit this number to a maximum of 10 in a given pixel, to avoid excessive contribution from \add{repeated} crossings due to particle gyration. Then, for each distribution of path lengths \add{of crossings within a given angular pixel}, 
\add{$s_p$ is found by an iterative fitting method that is binning-independent.} 

\begin{figure}
\centering
\includegraphics[width=.5\linewidth]{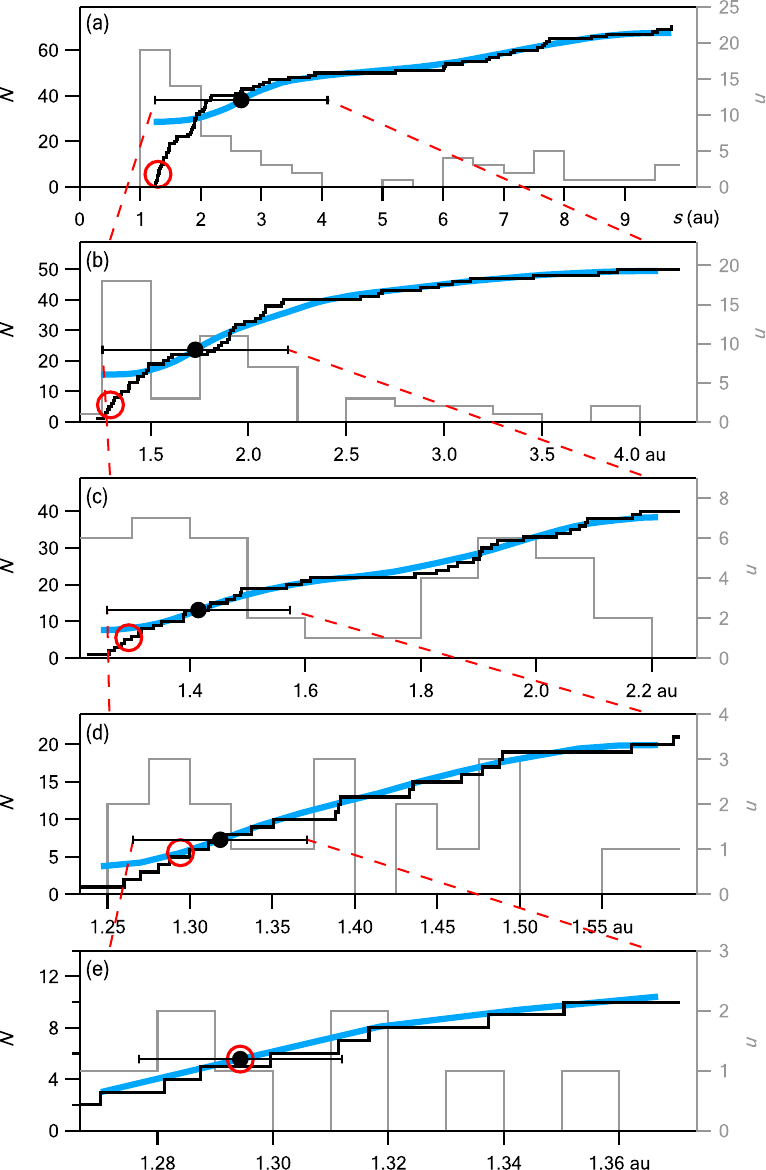}
\caption{\add{Iterative algorithm used to estimate $s_p$, the most probable (``peak'') value of path length, as shown for this example single-pixel distribution containing less than 100 particle crossings. (a)--(e) Each panel represents an iteration step $i$ and shows the same empirical cumulative distribution function $N(<s)$ (black), the smoothed distribution in this step (blue), and its point of maximum slope (black circle) as the estimate of $s_p$ with uncertainty $\xi_i/2$. For reference, the final estimate (red circle) is included on every panel. For comparison, a binned histogram $n(s)$ of the raw distribution with a bin width of one half of the axis tick spacing (grey) is also shown on each panel but is not used in the calculations.}}
\label{fig:peak_fit}
\end{figure}

\add{The peak fitting method is demonstrated in Figure \ref{fig:peak_fit}. First,}
an empirical cumulative distribution function \add{$N(<s)$ (black lines)} is constructed \add{from the crossings. 
Then in each step $i$, $N(<s)$ between $s_\mathrm{min}$ and $s_\mathrm{max}$\rev{, the minimum and maximum values of $s$,} is smoothed with a boxcar smoothing length of $\xi_i=(s_\mathrm{max}-s_\mathrm{min})/3$.
The smoothed cumulative distribution is indicated by blue lines in Figure \ref{fig:peak_fit}. 
The path length at its maximum slope is the estimate of $s_p$ at each iteration step $i$, which is taken to have uncertainty $\xi_i/2$. In the next step, the new $s_\mathrm{min}$ and $s_\mathrm{max}$ are set to  bracket that uncertainty interval (represented by red dashed lines), which implies that} $\xi_{i+1}=\xi_i/3$. 
The algorithm is terminated when the new interval width
is less than the distance from the Sun divided by 20.  

\add{Figure \ref{fig:peak_fit} also demonstrates our motivation for developing this binning-independent peak fitting technique.
The grey distributions are examples of binned histograms from the same underlying distribution, and for smaller bin widths the path length value where the histogram peaks is ambiguous and may fail to represent an optimal estimate of $s_p$.  
We find that our algorithm can robustly provide a reasonable ``compromise'' value of $s_p$ for a very wide variety of sparse distributions.
}
This algorithm is used both for global path length distributions and distributions within individual pixels.  
\add{Distributions of pixel peak path lengths are shown in Figure \ref{fig:pathlength_dist} (solid histograms).}

For spatial maps, in heliolongitude and heliolatitude, we plot only pixels with at least 70 crossings.
Note that some pixels are essentially inaccessible to outward-moving particles, a phenomenon associated with
observed ``dropouts'' of SEPs \citep{MazurEA00},
and may have very long peak path lengths associated with backscattering particles.
A detector at such a location would not register a strong SEP event.
In our maps we exclude pixels with peak path length $>1.05$ au 
(for $r_A=0.35$ au) or \add{$>$}1.7 au (for $r_B=1.0$ au), 
in order to focus on pixels with a significant population of outward-moving SEPs.
In maps of the number of particle crossings, we also exclude crossings of path length $>0.41$ au
(for $r_A$) or \add{$>$}1.7 au (for $r_B$); 
if long path lengths were not excluded, the number of crossings due to backscattering would increase without bound for longer simulation durations.

\section{Dependence of Path Lengths on Turbulence Amplitude and Composition}

We have explored the dependence of the path lengths of field lines and particles on the turbulence amplitude $b/B_0$ \rev{and} on the composition of 2D+slab turbulence.
We characterize the composition in terms of the fraction of fluctuation energy in the slab component, 
\begin{equation}
f_s = \frac{(b^\mathrm{slab})^2}{(b^\mathrm{slab})^2+(b^\mathrm{2D})^2}.
\end{equation}
This is particularly relevant to particle scattering, as only the slab component contributes to resonant pitch angle scattering, which is usually considered to be the dominant scattering process \citep{Jokipii66}.

In this section, we model gradual SEP events by initializing 5 million field lines and 5 million 1-MeV protons at the same set of (random) locations over the entire simulation domain at $r_0=0.1$ au. 
We record the path lengths of magnetic field lines and particle trajectories when they 
cross a surface at $r_A=0.35$ au or $r_B=1.0$ au.  
Figure \ref{fig:pathlength_dist} shows the global distributions of path lengths 
of field line (red) and particle (blue) crossings at $r_B$ for one set of parameter values and one slab representation.
The most probable (``peak'') path lengths are indicated by solid arrows.
As noted in Section 2, with no turbulence the field line path length would be $S=r$, and because we set the initial pitch angle to zero, the particle path length would also be $s=r$.
Then with the introduction of turbulence to the simulations, both types of path lengths can become substantially longer than $r$.

\begin{figure}
\centering
\includegraphics[width=.6\linewidth]{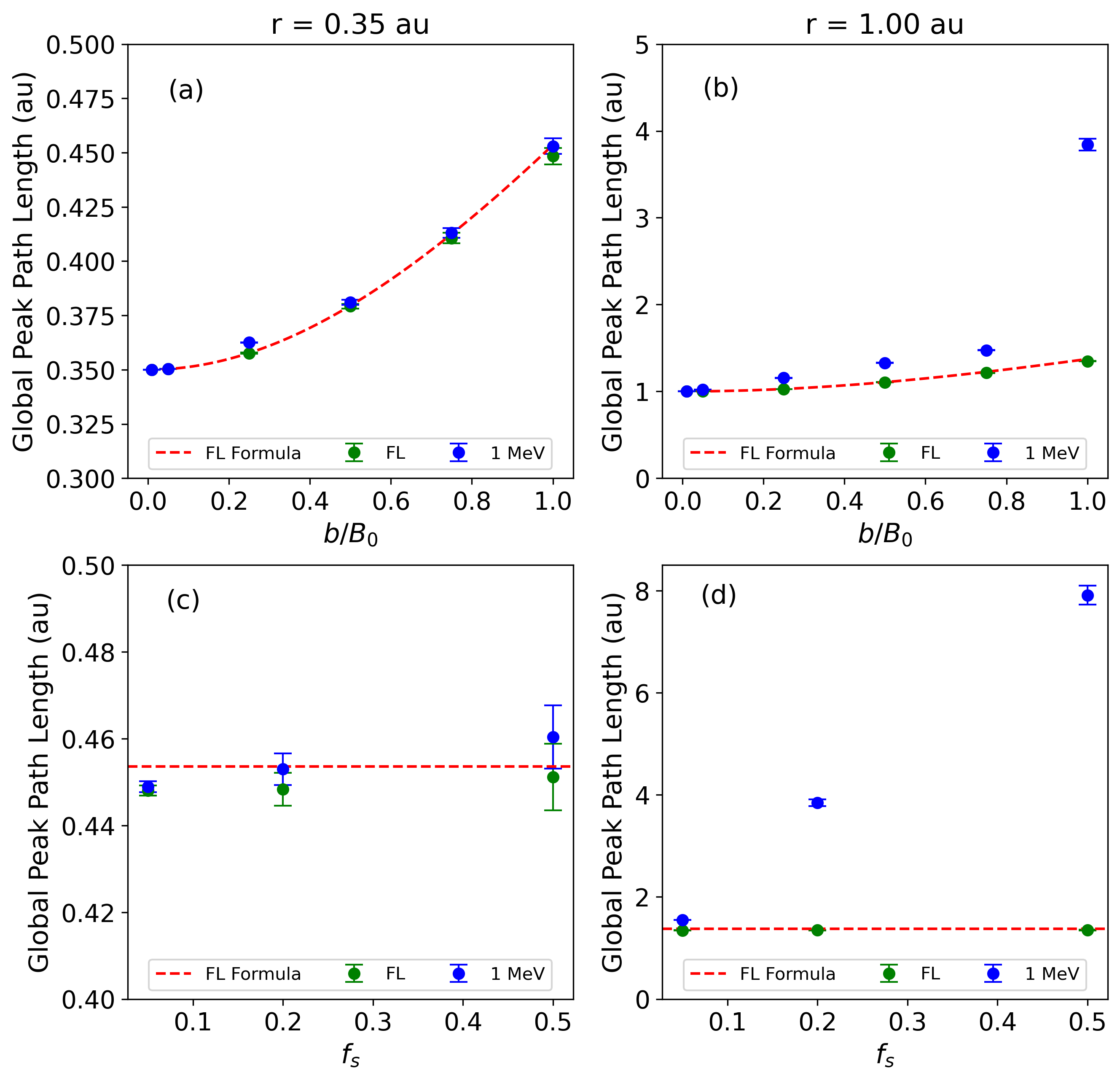} 
\caption{
Global peak path length for field lines (green points) and 1-MeV protons (blue points) as a function of turbulence amplitude $\dbob$ (with $f_s=0.2$) at (a) $r_A = 0.35$ au and (b) $r_B = 1.0$ au, and as a function of turbulence geometry as expressed by the slab fraction $f_s$ (with $\dbob=1$) at (c) $0.35$ au and (d) $1.0$ au, for a wide injection region. 
For field lines, the peak path length closely follows the simple approximation $S=r_0+r\sqrt{1+(\dbob)^2}(r-r_0)$ (dashed lines).
At $0.35$ au, 1-MeV proton path lengths closely follow those of field lines, though they are slightly longer for high $\dbob$. 
At $1.0$ au, particles experience strong scattering at high $\dbob$, and the global peak proton path length is much longer than that of field lines. 
Increasing the slab fraction enhances resonant particle scattering and leads to longer proton path lengths, especially at $1.0$ au, without significantly affecting the peak field line path length.} 
\label{fig:PPPL}

\end{figure}

We also want to characterize the widths of path length distributions; however, for distributions of particle path length $s$, some measures of the width are sensitive to the tail of the distribution at long $s$ that results from pitch angle scattering and spatial diffusion.  
In principle the distribution extends to infinite $s$, which of course is neither achieved in the simulations nor discernible in observations, nor is it of particular physical interest.
We therefore consider the difference between the path lengths at the peak of the distribution and at 10\% of the peak (indicated by dashed arrows in Figure \ref{fig:pathlength_dist}), which we call the ``rise,'' in order to characterize the 
\add{width of the rising phase}
of a distribution.  

To compute the rise, we use a histogram of path lengths for which the bin width is 0.01 times the peak path length (determined by the algorithm described in Section 2) minus distance from the Sun, or 0.001 au, whichever is smaller. 
Then the rise is determined as the difference between the path length value at the histogram peak and the preceding value at 10\% of the peak.

Furthermore, to better relate our results to spacecraft observations and to characterize spatial variation, we also determine path length distributions within $1^\circ$ by $1^\circ$ pixels.
We use the algorithm described in Section 2 to estimate the peak and rise for either global path length distributions or those for individual pixels
having at least 70 crossings.

Figure \ref{fig:PPPL} and Table 1 show how the path length distributions of field lines and 1-MeV protons depend on $b/B_0$ and $f_s$. 
In particular, Table 1 presents results both for the global distribution, in terms of its peak path length and rise, and for the variation among peak path lengths from individual $1^\circ\times1^\circ$ pixels, as characterized by their full width at half maximum (FWHM).
The FWHM is determined from a histogram with a bin width \add{chosen based on the peak path length value} as described above.
All results are based on five simulations, each using a different representation of slab turbulence. 
We report the mean value from the five simulations along with the uncertainty, calculated as the standard deviation among those simulations.

\begin{table*}
\caption{Statistics of path length distributions of field lines and 1-MeV protons at $r = 0.35$ au and $r= 1.0$ au for a \add{wide} injection \add{region}.\tablenotemark{a}}
\begin{center}
\begin{tabular}{ccclllllll}
\toprule
\multicolumn{10}{c}{\textit{Field Lines}} \\ Run & \multicolumn{1}{l}{\dbob} & $f_s$ & \multicolumn{3}{c}{$r$ = 0.35 au}    &                      & \multicolumn{3}{c}{$r$ = 1.0 au}     \\ \cline{4-6} \cline{8-10} 
& \multicolumn{1}{l}{}                        &                        & Peak Path & Rise & FWHM of & \multicolumn{1}{c}{} & Peak Path & Rise & FWHM of  \\                      & \multicolumn{1}{l}{}                        &                        & Length (au) & (au) & Pixel Peaks (au) & \multicolumn{1}{c}{} & Length (au) & (au) & Pixel Peaks (au) \\ \midrule
1	&	0.01	&	0.20	&	0.350	&	$<$ 0.01	&	$<$ 0.01	&	&		1.000	&	$<$ 0.01	&	$<$ 0.01	\\
2	&	0.05	&	0.20	&	0.350	&	$<$ 0.01	&	$<$ 0.01	&	&		1.001	&	$<$ 0.01	&	$<$ 0.01	\\
3	&	0.25	&	0.20	&	0.358	&	$<$ 0.01	&	$<$ 0.01	&	&		1.027	&	0.015	&	0.013	\\
4	&	0.50	&	0.20	&	0.379(1)	&	0.016	&	0.012(1)	&	&		1.105(1)	&	0.044	&	0.030(1)	\\
5	&	0.75	&	0.20	&	0.411(3)	&	0.028(2)	&	0.020(1)	&	&		1.214(3)	&	0.073(1)	&	0.047(3)	\\
6	&	1.00	&	0.05	&	0.448(1)	&	0.050(2)	&	0.030(1)	&	&		1.347(1)	&	0.139(1)	&	0.056(8)	\\
7	&	1.00	&	0.20	&	0.448(4)	&	0.039(3)	&	0.027(2)	&	&		1.348(4)	&	0.103(1)	&	0.055(4)	\\
8	&	1.00	&	0.50	&	0.451(8)	&	0.032(3)	&	0.023(3)	&	&		1.354(9)	&	0.080(2)	&	0.044(5)	\\

\midrule\addlinespace

\multicolumn{10}{c}{\textit{Protons of $E$ = 1 MeV}}                                                                                                                                                     \\ 
Run & \multicolumn{1}{l}{\dbob} & $f_s$ & \multicolumn{3}{c}{$r$ = 0.35 au}    & \multicolumn{1}{c}{} & \multicolumn{3}{c}{$r$ = 1.0 au}     \\ \cline{4-6} \cline{8-10} 
& \multicolumn{1}{l}{}                        &                        & Peak Path & Rise & FWHM of & \multicolumn{1}{c}{} & Peak Path & Rise & FWHM of  \\                      & \multicolumn{1}{l}{}                        &                        & Length (au) & (au) & Pixel Peaks (au) & \multicolumn{1}{c}{} & Length (au) & (au) & Pixel Peaks (au) \\
 \midrule
9	&	0.01	&	0.20	&	0.350	&	$<$ 0.01	&	$<$ 0.01	&	&		1.000	&	\add{$<$ 0.01}	&	$<$ 0.01	\\
10	&	0.05	&	0.20	&	0.350	&	$<$ 0.01	&	$<$ 0.01	&	&		1.021(1)	&	0.019(1)	&	$<$ 0.01	\\
11	&	0.25	&	0.20	&	0.363	&	0.010	&	0.007(2)	&	&		1.157(3)	&	0.093(2)	&	0.044(7)	\\
12	&	0.50	&	0.20	&	0.381(1)	&	0.016(1)	&	0.014	&	&		1.328(2)	&	0.124(1)	&	0.054(6)	\\
13	&	0.75	&	0.20	&	0.413(2)	&	0.028(2)	&	0.021	&	&		1.475(5)	&	0.136(3)	&	0.058(2)	\\
14	&	1.00	&	0.05	&	0.449(1)	&	0.050(2)	&	0.035(2)	&	&		1.549(2)	&	0.183(5)	&	0.09(1)	\\
15	&	1.00	&	0.20	&	0.453(4)	&	0.040(2)	&	0.029(2)	&	&		3.84(7)	&	2.33(8)	&	2.1(3)	\\
16	&	1.00	&	0.50	&	0.460(7)	&	0.033(2)	&	0.024(3)	&	&		$>$ 7	&	5.6(3)	&	3.8(7)	\\

\bottomrule
  
\end{tabular}
\end{center}
\tablenotetext{a}{Uncertainty in the final digit is indicated in parentheses, unless it is less than 1.}
\end{table*}

To validate our methodology for trajectory tracing in the presence of turbulence, we include the case of $b/B_0=0.01$ and verify the expected behavior in the turbulence-free limit.  
That is, the global peak path lengths of field lines and particles are indeed essentially equal to the radius $r$. 
\add{There} are very low values for the rise of the global path length distribution and the FWHM of the pixel peaks, 
\add{which we will subsequently refer to collectively as ``widths.''
This behavior approximately holds}
even for $b/B_0=0.05$.

With a stronger turbulence amplitude, $B$ increases significantly and hence we expect the path lengths to be longer.  
Based on Equations (\ref{eq:S}) and (\ref{eq:dSdr}), an approximate relationship could be
\begin{equation}
S \approx r_0 + \sqrt{1+b^2/B_0^2}\,(r-r_0),
\label{eq:approx}
\end{equation}
in terms of the mean squared magnetic fluctuation $b^2$.
This approximation is plotted with dashed lines in Figure \ref{fig:PPPL}, and indeed provides a good description of the dependence of the peak path length for field lines as a function of turbulence amplitude.
It is close (in fractional terms) to the even simpler expression $\sqrt{1+\left(b/B_0\right)^2}\,r$ in the limit of low $b/B_0$ or large $r/r_0$.

In the case of weak scattering, we might expect this formula to apply to particles as well, and Figure \ref{fig:PPPL} and Table 1 confirm that it usually provides a good approximation for low turbulence amplitude $b/B_0$ and for low slab fraction $f_s$.
Indeed, at $r_A=0.35$ au the peak path lengths of 1-MeV protons are similar to those of field lines in all cases.
At $r_B=1.0$ au, the proton peak path lengths are longer in all cases (except at $b/B_0=0.01$, where \add{the turbulence has almost no effect}), and they are much longer in the two cases with strong slab turbulence, i.e., $b/B_0=1$ and $f_s\geq0.2$, because of the effect of particle scattering \add{(Runs 15 and 16)}.
The same pattern is found for the rise of the global path length distributions and the FWHM of the pixel peak path lengths. 

Now let us consider the effect of the slab fraction $f_s$ for fixed turbulence amplitude ($b/B_0=1$). 
\add{As noted above, there are two basic types of behavior of the path length.
Scattering has a strong influence on the particles at 1 au for $b/B_0=1$ when there is}
strong slab turbulence ($f_s\geq0.2$, \add{Runs 15 and 16}), 
leading to very large values of the peaks and widths of path length distributions.
\add{The other type of behavior is found for particles at 0.35 au and for field lines, in which case}
the peaks of the global path length distributions  
remain nearly constant (within uncertainties) as a function of $f_s$.
\add{Then} we find that measures of the widths of the path length distributions {\it decrease} with increasing $f_s$ (though in some cases the decrease is not statistically significant within our simulation uncertainty). 
We can explain the reduced widths as follows: 
In the presence of only 2D fluctuations, field line trajectories would remain trapped along equipotential contours of $a(\varphi,\Lambda)$. 
This leads to spatial variation of path lengths; for example, field line and particle trajectories that are trapped along a contour with strong $|\mathbf{b^{\mathrm 2D}}|$ have particularly long path lengths.
However, an increasing slab fraction implies weaker trapping and therefore weaker spatial variation of path lengths.
This can explain the reduced rise of the global distribution and reduced FWHM of the pixel peak distribution with increasing $f_s$. 
\add{Another factor that could play a role is that the amplitude of 2D turbulence varies as $\sqrt{f_{2D}}=\sqrt{1-f_s}$ and decreases for increasing $f_s$, providing for less variation with $\varphi$ and $\Lambda$; however, that does not explain the noticeable difference between $f_s=0.05$ and 0.2 (for which $\sqrt{f_{2D}}=0.97$ and 0.89, respectively), which is better attributed to the trapping effect.}

\section{Dependence of Path Lengths on Turbulence Topology}

\begin{figure*}
     \centering
     \includegraphics[width=0.32\textwidth]{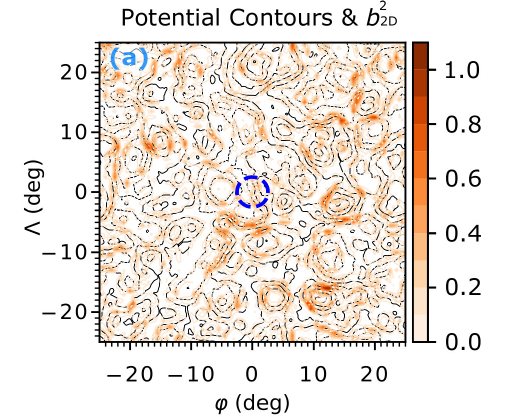}
     \includegraphics[width=0.32\textwidth]{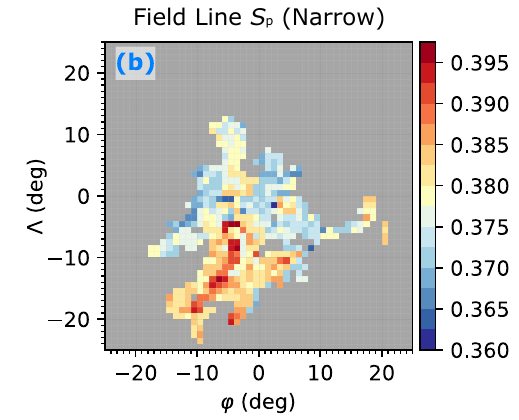}
     \includegraphics[width=0.32\textwidth]{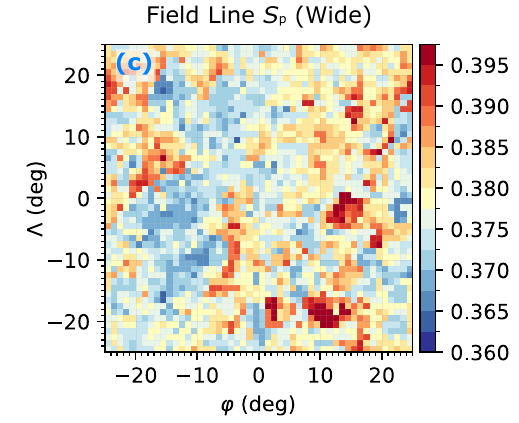}
     \includegraphics[width=0.32\textwidth] {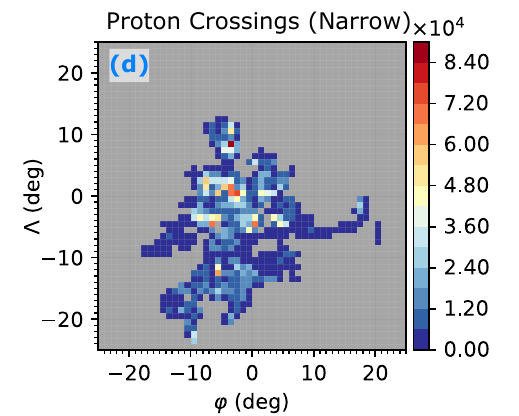}
\includegraphics[width=0.32\textwidth]{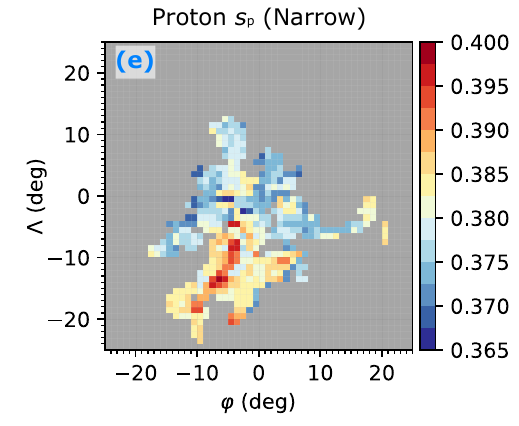}
\includegraphics[width=0.32\textwidth]{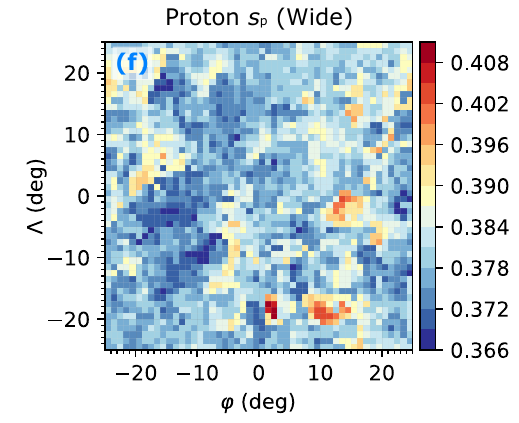}
\caption{
Maps in heliolongitude $\varphi$ and heliolatitude $\Lambda$ at radius $r_A=0.35$ au:  
(a) Contours of equal magnetic potential $a(\Lambda,\varphi)$, derived from a 2D MHD simulation of turbulence,
and resulting $|\mathbf{b^{\mathrm 2D}}|^2$ (color scale).
(b) Most probable (``peak'') path length in each pixel of $(\Lambda,\varphi)$ for field lines traced from a circle of radius 2.5$^\circ$ at $r_0=0.1$ au (blue dashed circle in (a)), to model the distribution of field lines connected to the narrow injection region of an impulsive solar event.
(c) Like (b), for field lines traced from all locations at $r_0=0.1$ au, to model the distribution of field lines connected to the wide injection region of a gradual solar event.
(d) Number of crossings at 0.35 au in each angular pixel for 1-MeV protons traced from the narrow injection region. 
(e) Peak length for crossings of 1-MeV protons traced from the narrow injection region.
(f) Like (e), but for the wide injection region. 
There are systematic variations in path length and arrival time that relate to the topology of 2D turbulence.
The spatial pattern of peak proton path lengths closely follows that for field lines, which in turn have longer path lengths in pixels where (a) indicates a strong 2D fluctuation field. 
}
        \label{fig:mapA}
\end{figure*}

Figure \ref{fig:mapA}(a) shows a contour map of the magnetic potential $a(\varphi,\Lambda)$, used to generate $\mathbf{b}^\mathrm{2D}(\varphi,\Lambda)=\nabla\times[a(\varphi,\Lambda)\mathbf{\hat{r}}]$, which then follows the equipotential contours.
For a mean field with only 2D fluctuations, the magnetic field lines would be 
forever trapped along such contours, representing an idealized flux tube (``spaghetti'') structure \citep{McCrackenNess66,MarianiEA73}.
In our simulations we also add a slab component, as suggested by observations of solar wind fluctuations \citep{MatthaeusEA90,BieberEA96}, so field lines experience only temporary topological trapping, giving rise to a dropout pattern \citep{RuffoloEA03}.
For the results in this section, we use a turbulence amplitude of $b/B_0=0.5$ and slab fraction $f_s=0.2$ \add{(Runs 4 and 12)}
\citep{BieberEA94}.

\begin{figure*}
     \centering
    \includegraphics[width=0.32\textwidth]{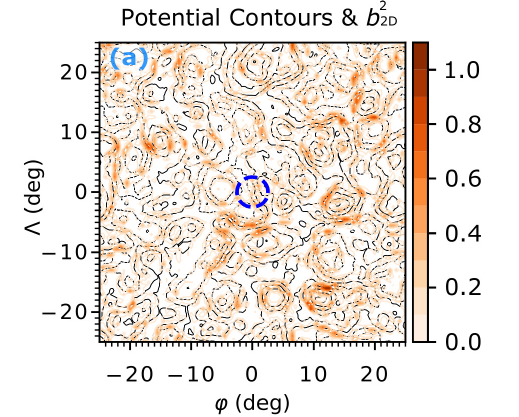}
    \includegraphics[width=0.32\textwidth]{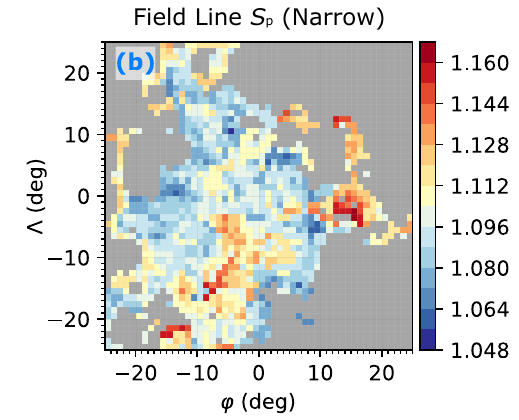}
      \includegraphics[width=0.32\textwidth]{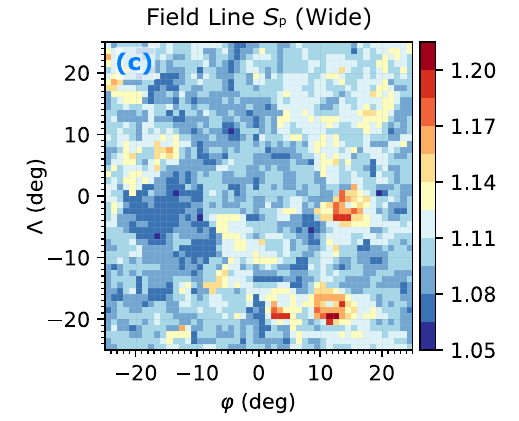}
     \includegraphics[width=0.32\textwidth]{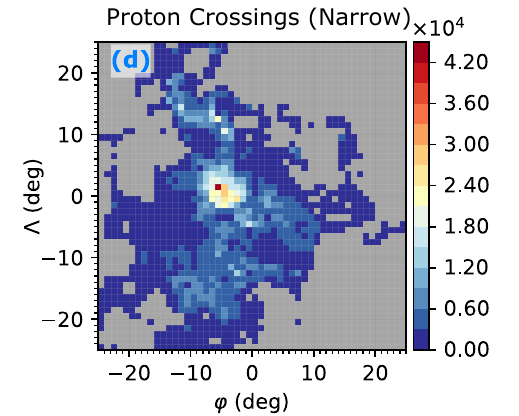}
\includegraphics[width=0.32\textwidth]{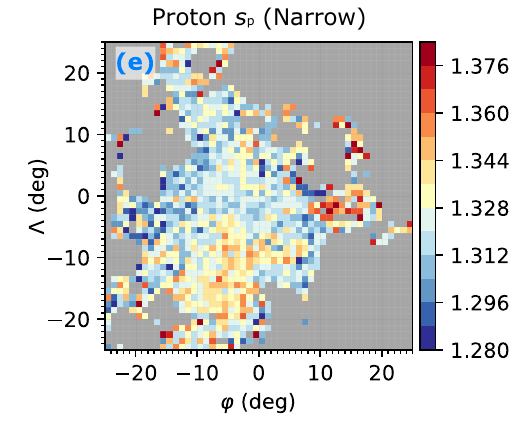}  
\includegraphics[width=0.32\textwidth]{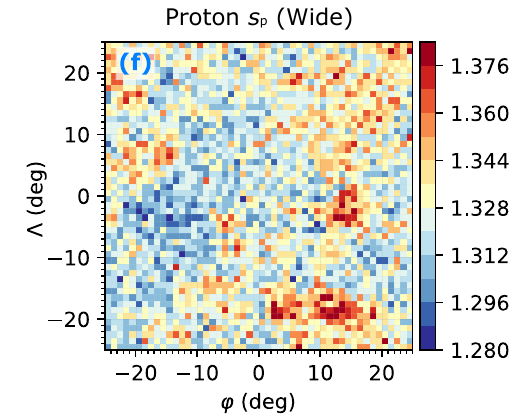}  
\caption{
Maps as in Figure \ref{fig:mapA}, but at $r_B=1.0$ au.
The spatial variation in path lengths again relates to the 2D fluctuation field at the point of observation.
}
        \label{fig:mapB}
\end{figure*}

Figures \ref{fig:mapA}(b-c) show the most probable (``peak'') path length $S_p$ of field lines in each pixel, for a narrow or wide injection region, respectively.
In either case, we see systematic variations that relate to the local $|\mathbf{b^{\mathrm 2D}}|^2$ {\it at the distance of observation}, as indicated by the red color scale (in units of $B_0^2$) in Figure \ref{fig:mapA}(a).
As noted earlier, the path length $S$ of a field line is the integral of $\mathrm{d}S/\mathrm{d}r=\sqrt{1+b^2(\mathbf{r})/B_0^2}$ {\it over its entire trajectory},
so apparently the local $|\mathbf{b^{\mathrm 2D}}|^2$ serves as a proxy of $b^2(\mathbf{r})$ over the trajectory \add{to some degree.  
There is a correlation coefficient of 0.35 between the $S_p$ data shown in Figure \ref{fig:mapA}(c) and the $|\mathbf{b^{\mathrm 2D}}|^2$ data shown in Figure \ref{fig:mapA}(a).}

We have also examined maps of the rise of the path length distribution within each pixel for narrow and wide injections.  
The patterns are qualitatively similar to those for peak path lengths, i.e., there is still an association with the local value of $|\mathbf{b^{\mathrm 2D}}|^2$.

Figure \ref{fig:mapA}(d) is a map of the number of 1-MeV proton crossings at 0.35 au in each pixel for the narrow injection region.
This is the irregular dropout pattern as previously reported from computer simulations \citep{GiacaloneEA00,RuffoloEA03,ZimbardoEA04,ChuychaiEA07,SeripienlertEA10,WangEA14,TooprakaiEA16,LaitinenEA23a},
which actually has finer structure that has been coarse-grained at the 1$^\circ$ level in this pixel map.
For a stronger turbulence amplitude $b/B_0$ we obtain stronger lateral spreading (and longer peak path lengths).
For a wide injection, as appropriate for modeling a gradual solar event, the distribution of crossings is much more uniform as has been noted previously \citep{GiacaloneEA00,RuffoloEA13} and is not shown in these Figures.

Figures \ref{fig:mapA}(e-f) show peak full-orbit path lengths of the protons arriving at each pixel, in analogy with Figures \ref{fig:mapA}(b-c) for field lines.  
There is a striking similarity between the patterns for protons and for field lines, as the proton path lengths to 0.35 au are only slightly longer than those of field lines.  
These results are for a proton energy of 1 MeV, and we obtained qualitatively similar results for an energy of 25 MeV.
Thus the spatial variation of SEP path lengths also bears the imprint of the 2D magnetic field pattern, which in turn relates to the flux-tube topology of the solar wind.

Figure \ref{fig:mapB} shows the corresponding results at $r_B=1.0$ au.  
The dropout pattern for a narrow (impulsive) injection has now spread to include farther heliolongitude $\varphi$ and heliolatitude $\Lambda$.  
Nevertheless, the maps of peak path lengths of field lines (Figures \ref{fig:mapB}(b-c)) remain qualitatively similar, retaining a clear correspondence with the local strength of $|\mathbf{b^{\mathrm 2D}}|^2$ (Figure \ref{fig:mapB}(a)).  

Interestingly, Figure \ref{fig:mapB}(d) indicates a more centrally concentrated distribution of 1-MeV proton crossings at the distance of 1 au than was found at 0.35 au (Figure \ref{fig:mapA}(d)).
\add{However, this is not a general result; in simulations for other slab representations, the distribution at 1 au is sometimes less centrally concentrated}.
\cite{ChhiberEA21apj} provided a theoretical framework to describe the \add{expected} angular width of such a \add{distribution}.

Finally, maps of peak path lengths for 1-MeV protons at 1.0 au are shown in Figures \ref{fig:mapB}(e-f) for a narrow and wide injection, respectively.  
At this distance from the Sun, pitch angle scattering has a stronger effect and makes these full-orbit particle path lengths significantly longer than those of the corresponding field lines.
The effect of random scattering also leads to noticeable random fluctuations in the maps.  
Nevertheless, the same qualitative pattern persists, in association with the local $|\mathbf{b^{\mathrm 2D}}|^2$.
We note further that for a narrow injection region, 
\add{our results for different slab representations do not}
indicate a 
\add{systematic} dependence of the peak path length or rise of field lines or particles on angular distance from the injection region, though in the case of particles at 1.0 au there is stronger statistical scatter at larger angular distance in association with the lower number of particle crossings and the randomizing effect of pitch angle scattering.

\section{Discussion}
\label{Sec:disc}

In this work, an overall theme is the effect of temporary topological trapping of magnetic field lines by the flux-tube structure of 2D turbulence \citep{RuffoloEA03}.  
Such structure is well documented from solar wind magnetic field and plasma observations \citep[e.g.,][and references therein]{Borovsky08}, yet field lines should not be expected to be permanently confined within flux tubes as noted by \citet{MatthaeusEA95}.
In the idealized 2D+slab model of magnetic fluctuations as employed here, ``islands'' defined by closed equipotential contours of the 2D turbulence correspond to flux surfaces in three dimensions.  
In nature these surfaces are certainly not at the same angular coordinates $\varphi$ and $\Lambda$ for all distances from the Sun; yet there is evidence that they do maintain their integrity to a distance of 1 au because of the observation of SEP dropouts \citep[e.g.,][]{MazurEA00,HoEA22}, which can naturally be explained by temporary topological trapping of field lines \citep{RuffoloEA03} in the pre-existing 2D+slab model.  

Indeed the persistence of certain magnetic field properties after transport over large radial distances even in the presence of turbulence may have application in other areas of heliospheric physics.  
For example 
the organization of the photospheric magnetic field at supergranulation scales, with substructure at granulation scales, might be thought of crudely analogous to the island structure of the 2D turbulence that we employ in  the present model.  
It is noteworthy that PSP observations \citep{BaleEA21} 
have established that at 10 to 20 solar radii one finds clusters of magnetic activity that map well, in a statistical sense, back to the photospheric supergranulation.  
Based on our findings, this kind of imprint of magnetic properties might be a natural expectation, even if, due to effects of the turbulence, the suggestion is of a statistical association rather than a one-to-one mapping. 

The most probable (peak) path length of either field lines or particles, as a function of heliolongitude and heliolatitude, is found to be associated with the {\it local} value of $|\mathbf{b^{\mathrm 2D}}|^2$, even though the path length of a field line represents the integral of $B/B_0$ along its entire trajectory.
The slab field is nearly independent of $(\varphi,\Lambda)$,
so the path length variation pattern indicates that $|\mathbf{b^{\mathrm 2D}}|^2$ at the distance of observation serves as a proxy of its value throughout the trajectories.
In some regions, this reflects a topological trapping effect of confinement of field lines to islands of the 2D turbulence, 
as in the case of several islands with noticeably long path lengths of field lines in Figures \ref{fig:mapA}(c) and \ref{fig:mapB}(c) and of particles in Figures \ref{fig:mapA}(f) and \ref{fig:mapB}(f).
In other regions, the $(\varphi,\Lambda)$ trajectories along the 2D equipotential contours, when not significantly perturbed by the slab turbulence, traverse long regions of strong $|\mathbf{b^{\mathrm 2D}}|^2$, such as the region with $\varphi$ near zero and $\Lambda$ ranging from about $-5^\circ$ to $-20^\circ$, where there is a particularly pronounced enhancement in path lengths in Figures \ref{fig:mapA}(b), \ref{fig:mapA}(e), and \ref{fig:mapB}(b). 
This proxy effect implies a possible measurable association between SEP pathlengths and local magnetic measurements.
The magnetic helicity as measured by a single spacecraft \citep{MatthaeusJGR82} is a useful quantity to indicate a strongly trapping flux tube \citep{PecoraEA21}, and indeed an association has been found between structures with strong magnetic helicity and SEP dropouts \citep{TrenchiEA13}.

The peak path length of field lines is generally well described by Equation (\ref{eq:approx}) or in cases of low $b/B_0$ or large $r/r_0$, by the even simpler expression 
$S \approx \sqrt{1+\left(b/B_0\right)^2}\,r$, where $b$ refers to the rms turbulence amplitude.
As shown in Figure \ref{fig:pathlength_dist}, particles generally have longer path lengths than \add{field lines} as a result of pitch angle scattering.  
We find a much greater increase in path lengths of particles relative to field lines at a distance of 1.0 au than at 0.35 au.
Close to the Sun, the effect of adiabatic focusing (magnetic mirroring) in the interplanetary magnetic field is quite strong and the pitch angle distribution is approximately described by an equilibrium between focusing and scattering \citep{RuffoloK95}, with most particles moving outward.
The greater effect of scattering on the particle path lengths at greater distance from the Sun  relates to the reduced effect of focusing and increased scattering after particles travel a greater number of mean free paths. 
Particle path lengths are greatly enhanced in cases of strong scattering, i.e., for higher turbulence amplitude $b/B_0$ or slab fraction $f_s$, because the slab component of turbulence is responsible for resonant pitch angle scattering.
For other cases, particle path lengths are not much greater than field line path lengths and share a similar dependence on turbulence parameters.
While a greater $b/B_0$ leads to longer path lengths according to Equation (\ref{eq:approx}), and also to longer rise of global path length distributions and FWHM of pixel peak path lengths, increasing $f_s$ actually tends to reduce the rise and FWHM in these cases, because a stronger slab fraction causes more rapid escape from turbulent trapping of field lines and path length distributions that are more spatially homogeneous. 

We note that the recent study of magnetic field line path lengths by \citet{LaitinenEA23a} also traced magnetic field line trajectories in representations of 2D+slab turbulence, and used a large-scale magnetic field along a Parker spiral.  
They showed results for the path length distribution as a function of heliolongitude and indeed found fluctuations in this distribution.
The turbulence energy used in that work, $b^2/B_0^2\approx0.6$ at $r=1$ au, is comparable to one of the cases in our Table 1, with $b/B_0=0.75$. 
The peak field line path lengths are similar when accounting for the length along the Parker spiral mean field in that work (1.17 au) compared with 1.0 au along the radial mean field in the present work.
Their work confirms that the peak field line path length is reasonably well described by the simple expression of \citet{ChhiberEA21aa}, which corresponds to Equation (\ref{eq:approx}) in the present work.
Their corresponding results for 100-MeV proton path lengths \citep{LaitinenEA23b} are also comparable to the results in our Table 1 for 1-MeV protons with $b/B_0=0.75$, when accounting for the 0.17 au path length difference along the mean field lines.

In a given simulation, we have used constant values of $b/B_0$, $f_s$, and the spectral index of the inertial range of the turbulent power spectrum.  
This can be justified for $b/B_0$ because the radial dependence over $r<1$ au is weak \citep{Chhiber22} and for $f_s$ because we lack a precise knowledge of how it varies with $r$.  
Note that while some studies have found $f_s\approx0.2$ near 1 au \citep{BieberEA94,BieberEA96}, \citet{BandyopadhyayM21} analyzed PSP data over 0.13 au $\leq r \leq 0.25$ au and concluded that $f_s\sim0.5$ in this region.
These studies imply that $f_s$ varies with distance from the Sun.  
Here we view that our constant value of $f_s$ describes a typical value over the distance considered, and one might expect higher values to apply over shorter distances from the Sun.
For the spectral index of the components of magnetic turbulence (specifically, for prescription of the slab turbulence and the input to the 2D MHD dynamical model), we have used a constant value of $-5/3$ as in Kolmogorov theory.
A recent analysis of PSP data showed that the spectral index of the inertial range of solar wind turbulence varies with $r$ from $-5/3$ near $r=1$ au to $-3/2$ closer to the Sun \citep{ChenEA20}.
This difference should be negligible in the context of our study, because our results show that peak pathlengths of field lines at a given $r$ are well described by Equation (\ref{eq:approx}), which depends only on the turbulence amplitude, and the effects of turbulence topology, which depend on the large-scale flux-tube structure of the 2D turbulence component, not the details of the inertial range.

Path length variations include the spatial variations shown in Figures \ref{fig:mapA} and \ref{fig:mapB} as well as variation in terms of a distribution of path lengths of field lines and particles arriving at a location of interest, e.g., the location of a detector.  
Indeed, an observed time-intensity profile of SEPs from a near-solar source represents the distribution in path length $s$, converted to travel time $t=s/v$ and convoluted with the injection profile.
In addition, SEP modeling should allow for field line path lengths $S$ that are longer than that along the mean field (which in this work would be $S=r$) as previously described by theory and simulations \citep{ChhiberEA21aa}, with some uncertainty due to the variation with heliolongitude and heliolatitude as described here, as well as unknown parameters of interplanetary turbulence.
Therefore, the path length of the guiding field lines in a transport model could be treated as an adjustable parameter.
As noted earlier, it is not appropriate to identify a field line path length $S$ with the typical path length $s$ of particles near onset as determined by velocity dispersion analysis, as the latter is greater due to interplanetary scattering.

We find differences between the global path length distribution and the path length distributions at individual $1^\circ\times1^\circ$ pixels in heliolongitude and heliolatitude.  
The global distribution is broader because it incorporates the distributions of all pixels.  
If an SEP transport model does not consider a representation of turbulence, but rather considers turbulence to impart random kicks to particles, then it is possible that the global distribution is effectively being used, rather than the narrower and more physically appropriate distribution for a detector at a specific location in the turbulence topology.
In the forward modeling approach, this artificially broadens the expected SEP time-intensity profiles.
In the backward modeling approach, this can affect the fitting results, tending to favor a shorter duration of injection (because the modeling artifically broadens the time profile) and/or a longer mean free path of interplanetary scattering, or in some cases makes it impossible to obtain a good fit to the data.

Another effect of the spatial variation of the path lengths on SEP transport is that during the course of an SEP event observation, the dropout pattern can be convected with the solar wind flow past the spacecraft; indeed this is generally considered to explain why the spatial pattern of magnetic connectivity can result in sharp time variations of dropouts in impulsive SEP events \citep{MazurEA00,GiacaloneEA00}. 
Furthermore, this effect of changing magnetic connectivity causing sharp time variations has also been observed in some gradual SEP events
\citep{NemzekEA94,RuffoloEA06}.
While this could reflect the different density of SEPs along different flux tubes, it is also affected by spatial variation in the path length distribution, which can suddenly shift the time-intensity profile of one location (one ``pixel'' in our work) relative to that in a neighboring location.
These effects may not be considered in existing models of SEP transport, especially when effects of turbulence are treated as stochastic processes, not taking into account the flux-tube topology and temporary trapping as expected in representations of solar wind turbulence.

\begin{acknowledgments}
This research has been supported in Thailand by Thailand Science Research and Innovation (RTA6280002), by the National Science and Technology Development Agency (NSTDA) and National Research Council of Thailand (NRCT): High-Potential Research Team Grant Program (N42A650868), and from the NSRF via the Program Management Unit for Human Resources \& Institutional Development, Research and Innovation (B37G660015).
It was also supported by the {\it Parker Solar Probe} mission under the 
 ISOIS project 
 (contract NNN06AA01C) and a subcontract 
 to University of Delaware from
 Princeton University (SUB0000165)
 and by the IMAP project through a Princeton subcontract (SUB0000317).
 Additional support is acknowledged from NASA under the 
 LWS program  (NNX17AB79G, 80NSSC22K1020), the HSR program (80NSSC18K1210 \& 80NSSC18K1648), and the Helio PSP-GI program (80NSSC21K1765).
\end{acknowledgments}

\newcommand{\jgra}{\jgr\ (Space Phys.)}

\end{document}